\documentclass[bibyear]{aa}  

\usepackage{graphicx}
\usepackage{txfonts}

\usepackage{comment}
\usepackage{caption} 
\usepackage{natbib}
\bibpunct{(}{)}{;}{a}{}{,} 

\usepackage[colorlinks]{hyperref}
\hypersetup{
    colorlinks = True,
    citecolor = blue
}

\usepackage{acronym}
\acrodef{BH}[BH]{black hole}

\newcommand{\orcit}[1]{\protect\href{https://orcid.org/#1}{\protect\includegraphics[width=8pt]{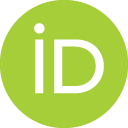}}}

\newcommand{\modref}[1]{{#1}}
\newcommand{\modlref}[1]{{#1}}

\begin{document}

   \title{The boring history of Gaia BH3 from isolated binary evolution}

   \subtitle{}

   \author{Giuliano Iorio
           \orcit{0000-0003-0293-503X}\inst{1,2,3}\thanks{\href{mailto:giuliano.iorio.astro@gmail.com}{giuliano.iorio.astro@gmail.com}},
          Stefano Torniamenti \orcit{0000-0002-9499-1022}\inst{4,2}\thanks{\href{mailto:stefano.torniamenti@uni-heidelberg.de}{stefano.torniamenti@uni-heidelberg.de}},
          Michela Mapelli\orcit{0000-0001-8799-2548}\inst{4,1,2,3}\thanks{\href{mailto:mapelli@uni-heidelberg.de}{mapelli@uni-heidelberg.de}},
          Marco Dall'Amico\orcit{0000-0003-0757-8334}\inst{1,2},
          Alessandro A. Trani\orcit{0000-0001-5371-3432}\inst{5},
          Sara Rastello\orcit{0000-0002-5699-5516}\inst{6},
          Cecilia Sgalletta\orcit{0009-0003-7951-4820}\inst{2,7},
          Stefano Rinaldi \orcit{0000-0001-5799-4155} \inst{1,4},
          Guglielmo Costa\orcit{0000-0002-6213-6988}\inst{8,3},
          Bera A. Dahl-Lahtinen\inst{1},
          Gast\'on J. Escobar\orcit{0000-0002-4007-7585}\inst{1,2},
          Erika Korb\orcit{0009-0007-5949-9757}\inst{1,2,4},
          M. Paola  Vaccaro\orcit{0000-0003-3776-9246}\inst{4},
          Elena Lacchin\orcit{0000-0001-9936-0126}\inst{1,2,4,9},
          Benedetta Mestichelli\orcit{0009-0002-1705-4729}\inst{4,10,11},
          Ugo N. Di Carlo\orcit{0000-0003-2654-5239}\inst{7},
          Mario Spera\orcit{0000-0003-0930-6930}\inst{7},
          Manuel Arca Sedda\orcit{0000-0002-3987-0519}\inst{10,11,12}
          }

   \institute{
   Physics and Astronomy Department Galileo Galilei, University of Padova, Vicolo dell'Osservatorio 3, I--35122, Padova, Italy 
   \and
   INFN - Padova, Via Marzolo 8, I--35131 Padova, Italy
   \and
    INAF – Padova, Vicolo dell’Osservatorio 5, I-35122 Padova, Italy
   \and
   Institut f{\"u}r Theoretische Astrophysik, ZAH, Universit{\"a}t Heidelberg, Albert-Ueberle-Stra{\ss}e 2, D-69120, Heidelberg, Germany
    \and
    Niels Bohr International Academy, Niels Bohr Institute, Blegdamsvej 17, 2100 Copenhagen, Denmark
    \and
    Departament de F\'isica Qu\`antica i Astrof\'isica, Institut de Ci\`encies del Cosmos, Universitat de Barcelona, Mart\'i i Franqu\`es 1, E-08028 Barcelona, Spain
    \and
    SISSA, via Bonomea 365, I-34136 Trieste, Italy
    \and
    Univ Lyon, Univ Lyon1, ENS de Lyon, CNRS, Centre de Recherche Astrophysique de Lyon UMR5574, F-69230 Saint-Genis-Laval, France
    \and 
    INAF, Osservatorio Astronomico di Bologna, Via Gobetti 93/3, I-40129 Bologna, Italy 
    \and
    Gran Sasso Science Institute, Via F. Crispi 7, I-67100 L'Aquila, Italy
    \and
    INFN - Laboratori Nazionali del Gran Sasso, I-67100 L'Aquila, Italy
    \and
    INAF - Osservatorio Astronomico d’Abruzzo, Via Mentore Maggini, s.n.c., I-64100 Teramo, Italy\\}

   \date{Received 26-04-2024; accepted 31-07-2024}
 
  \abstract{{\it Gaia}~BH3 is the first observed dormant \ac{BH} with a mass of $\approx{30}$ M$_\odot$, and it represents the first confirmation that such massive \ac{BH}s are associated with metal-poor stars. Here, we explore the isolated binary formation channel for {\it Gaia}~BH3, focusing on the old and metal-poor stellar population of the Milky Way halo. We used the {\sc mist} stellar models and
  our open-source population synthesis code {\sc sevn} to evolve \modref{$5.6 \times 10^8$} binaries, exploring \modref{20}  sets of parameters that encompass different natal kicks, metallicities, common envelope \modref{efficiencies and binding energies}, and models for the Roche-lobe overflow. 
  We find that systems such as {\it Gaia}~BH3 form preferentially from binaries initially composed of a massive star ($40-60$ M$_\odot$) and a low-mass companion ($<1$ M$_\odot$) in a wide ($P>10^3$ days) and eccentric orbit ($e>0.6$). Such progenitor binary stars do not undergo any Roche-lobe overflow episode during their entire evolution, so the final orbital properties of the BH-star system are essentially determined at the core collapse of the primary star.  Low natal kicks  ($\lesssim$ 10~km/s) significantly favour the formation of {\it Gaia}~BH3-like systems, but high velocity kicks up to $\approx 220$ km/s are also allowed. We estimated the formation efficiency for {\it Gaia}~BH3-like systems in old ($t>10$ Gyr) and metal-poor ($Z<0.01$) populations to be $\sim 4 \times 10^{-8}$ M$_\odot^{-1}$ (for our fiducial model), representing  $\sim 3\%$ of the whole simulated \modlref{BH-star} population. We expect up to $\approx 4000$ \modlref{BH-star} systems in the Galactic halo formed through isolated evolution, of which $\approx 100$ are compatible with {\it Gaia}~BH3. 
  {\it Gaia}~BH3-like systems represent a common product of isolated binary evolution at low metallicity ($Z<0.01$), but given the steep density profile of the Galactic halo, we do not expect more than one 
  at the observed distance of {\it Gaia}~BH3. 
  \modref{Our models show that even if it was born inside a stellar cluster, {\it Gaia}~BH3 is
compatible with a primordial binary star that escaped from its parent 
cluster without experiencing significant dynamical interactions.} 
} 


   \authorrunning{G. Iorio et al.}

   \maketitle

\section{Introduction}

{\it Gaia}  data \citep{GaiaDR3} offer a unique opportunity to search for dormant black holes (BHs), that is X-ray and radio-quiet BHs that are members of a binary system with a star  \citep{guseinov1966,trimble1969}. {\it Gaia}~BH3, the third dormant BH retrieved from these data, has a mass of $\approx{33}$~M$_\odot$ and is member of a loose eccentric binary system (orbital period $P\approx{4000}$ days, eccentricity $e\approx{0.72}$) that includes a $\approx{0.76}$~M$_\odot$ very metal-poor giant star ($[\textrm{Fe}/\textrm{H}]=~-2.56\pm{0.12}$ \modlref{obtained} from the spectrum of the Ultraviolet and Visual Echelle Spectrograph; \citealt{GaiaBH3}).

{\it Gaia}~BH3 is the first-ever discovered dormant BH with a mass $\gtrsim{}30$~M$_\odot$ and the first stellar BH with such a high mass unequivocally observed in the Milky Way. The other two {\it Gaia} BHs (BH1 and BH2) both have a mass of $\approx{10}$ M$_\odot$ \citep{elbadry2023a,elbadry2023b,chakrabarti2023,tanikawa2024}. All the other dormant BH candidates \citep{casares2014,ribo2017,giesers2018,giesers2019,thompson2019,shenar2022a,shenar2022b,mahy2022,lennon2022,saracino2022,saracino2023} and the BHs in X-ray binary systems with a dynamical measurement have a mass between a few and $\approx{20}$~M$_\odot$ \citep{oezel2010,millerjones2021}.

Stellar BHs with a mass $\gtrsim{30}$ M$_\odot$ have been observed for the first time with gravitational waves \citep{abbottGW150914,abbottGW150914astro,abbottGWTC2,abbottGWTC2.1,abbottGWTC3}. According to the main formation scenario, such massive BHs are expected to form from metal-poor stars, which do not lose much mass by stellar winds and collapse to a BH directly \citep{heger2002,mapelli2009,mapelli2010,mapelli2013,zampieri2009,belczynski2010,fryer2012,ziosi14,spera2015}. Gravitational wave data have proven the existence of such `oversized' BHs, but they do not carry any information about the progenitor's metallicity because no electromagnetic counterpart has been observed. Hence, {\it Gaia}~BH3 is the first direct confirmation that BHs with a mass in excess of 30~M$_\odot$ are associated with metal-poor stellar populations, thus supporting the main theoretical scenario.

For both its chemical composition and Galactic orbit, {\it Gaia}~BH3 is associated with the ED--2 stream, which is composed of stars that are at least 12 Gyr old \citep{balbinot2024}. Thus, this is the first BH unambiguously associated with a disrupted star cluster with a mass of $2\times{}10^3-4.2\times{}10^4$~M$_\odot$ \citep{balbinot2024}.  Preliminary numerical models have shown that {\it Gaia}~BH3 might have assembled dynamically from a low-mass star and a BH that did not evolve in the same binary star system \citep[][]{marinpina2024}.

While the association of {\it Gaia}~BH3 with the ED--2 stream is indisputable, in this work we show that this system is compatible with the nearly unperturbed evolution of a primordial binary star. Our population-synthesis simulations 
show that we do not need to invoke a dynamical exchange or a dynamical capture to explain its orbital properties. 

{\it Gaia}~BH3 is very different from both {\it Gaia} BH1 and {\it Gaia} BH2, which challenge our models of binary evolution (\citealt{elbadry2023a,elbadry2023b,chakrabarti2023,tanikawa2023bh,rastello2023,dicarlo2024,generozov2024}, but see \citealt{kotko2024}). The wide orbital separation of {\it Gaia}~BH3 does not require evolution via Roche-lobe overflow or a common envelope. 
Rather, it is consistent with a wide system that never went through mass transfer and survived the formation of the BH because 
it happened via an \modref{almost} direct collapse with a  mild natal kick. 

In this work, we show that the measured orbital eccentricity helps constrain the phase of the binary at the time of the formation of the BH. 
This paper is organised as follows. We describe our methodology in Sect.~\ref{sec:method},  present the main results in Sect.~\ref{sec:results}, and discuss the implications in Sect.~\ref{sec:discussion}. Section~\ref{sec:summary} provides a summary of our main takeaways.

\section{Methods} \label{sec:method}

In this work, we compare the observational features of the {\it Gaia}~BH3 system with a synthetic population of stellar binaries produced with the Stellar EVolution for N-body (\textsc{sevn}) code\footnote{In this work, we used the {\sc sevn} version V 2.10.1 (commit \href{https://gitlab.com/sevncodes/sevn/-/tree/a4753f1177aab076da2641afc497a06006b30267}{a4753f11}) publicly available at the GitLab repository \url{https://gitlab.com/sevncodes/sevn}.}  \citep{spera2017,spera2019,mapelli2020,iorio2023}. 
In {\sc sevn}, the evolution of single stars is estimated by interpolating on the fly across pre-calculated stellar evolution tracks stored in custom lookup tables. The models presented in this study were computed using the {\sc mist} stellar tracks by \cite{choi2016}. 
 We used {\sc trackcruncher}\footnote{The version of the code we used to produce the tables is available at \url{https://gitlab.com/sevncodes/trackcruncher} (commit \href{https://gitlab.com/sevncodes/trackcruncher/-/tree/08273372c1f43cc545bae2e06d618e81962f8419}{0827337}).} \citep{iorio2023} to produce the {\sc sevn} tables from the {\sc mist} stellar tracks. We produced eight sets of tables from metallicity $Z=1.4 \times 10^{-5}$ to $4.5 \times 10^{-2}$; each set contains stellar tracks from 0.71 M$_\odot$ to 150 M$_\odot$.

In {\sc sevn}, binary interactions are described through analytical and semi-analytical prescriptions, 
encompassing stable mass transfer by Roche-lobe overflow and winds, common envelope evolution, angular momentum dissipation by magnetic braking, tidal interactions, orbital decay by gravitational-wave emission, dynamical hardening, chemically homogeneous evolution, and stellar mergers. In our models, we treated tidal forces with the analytical formalism developed by \cite{hut1981} and as implemented by \cite{hurley2002}. For more details, we refer to \cite{iorio2023}.

\subsection{Initial conditions}\label{sec:ic}
We generated the initial conditions for  $2 \times 10^7$ binary systems by extracting the zero-age main-sequence mass of the primary (more massive) star, $m_1$, from the Kroupa initial mass function 
\citep{kroupa2001} from 5  to 150 M$_\odot$. The mass of the secondary (less massive) star was set from the binary mass ratio $q$ following the distribution PDF$(q) \propto q^{-0.1}$ with $q\in\left[ 0.71\,{}\textrm{M}_\odot/m_1,1\right]$.
We obtained the initial orbital period ($P$)  and eccentricity ($e$) from the following distributions: PDF$\left( \log P/\mathrm{days}\right) \propto \left(\log P\right)^{-0.55}$, PDF$(e) \propto e^{-0.42}$.  These distributions of mass ratios, orbital periods, and eccentricities \modlref{represent the best fits to} observations of massive binary stars in Galactic young clusters \citep{sana2012}. The periods range from 1.4 days to 866 years, while the eccentricities range from 0 to $e_\mathrm{max}=1-\left(P/2 \ \mathrm{days}\right)^{-2/3}$, according to \cite{moe2017}. 
The total mass of the simulated population is $M_{\rm{sim}}=4.2\times10^8 \ \mathrm{M}_\odot$, corresponding to a mass of 
$M_{\rm{tot}}=1.4\times10^9 \ \mathrm{M}_\odot$ \modref{(roughly equivalent to the mass of the Galactic stellar halo; see \citealt{deason2019})} when accounting for the fact that we did not simulate low-mass primary stars and systems with secondary stars less massive than 0.71 M$_\odot$.\footnote{We assumed a parent population with primary masses $m_1$ between 0.08 M$_\odot$ and 150 M$_\odot$ and mass ratios in the range $q\in [0.08\,{}\textrm{M}_\odot/m_1,1]$.} 
We used {\sc sevn} to evolve all the generated binaries from the zero age main sequence up to the moment in which the two stars are compact remnants (white dwarfs, neutron stars, \ac{BH}s).

\subsection{Fiducial model} \label{sec:fiducial}

   \begin{figure}
   \centering
   \includegraphics[width=\hsize]{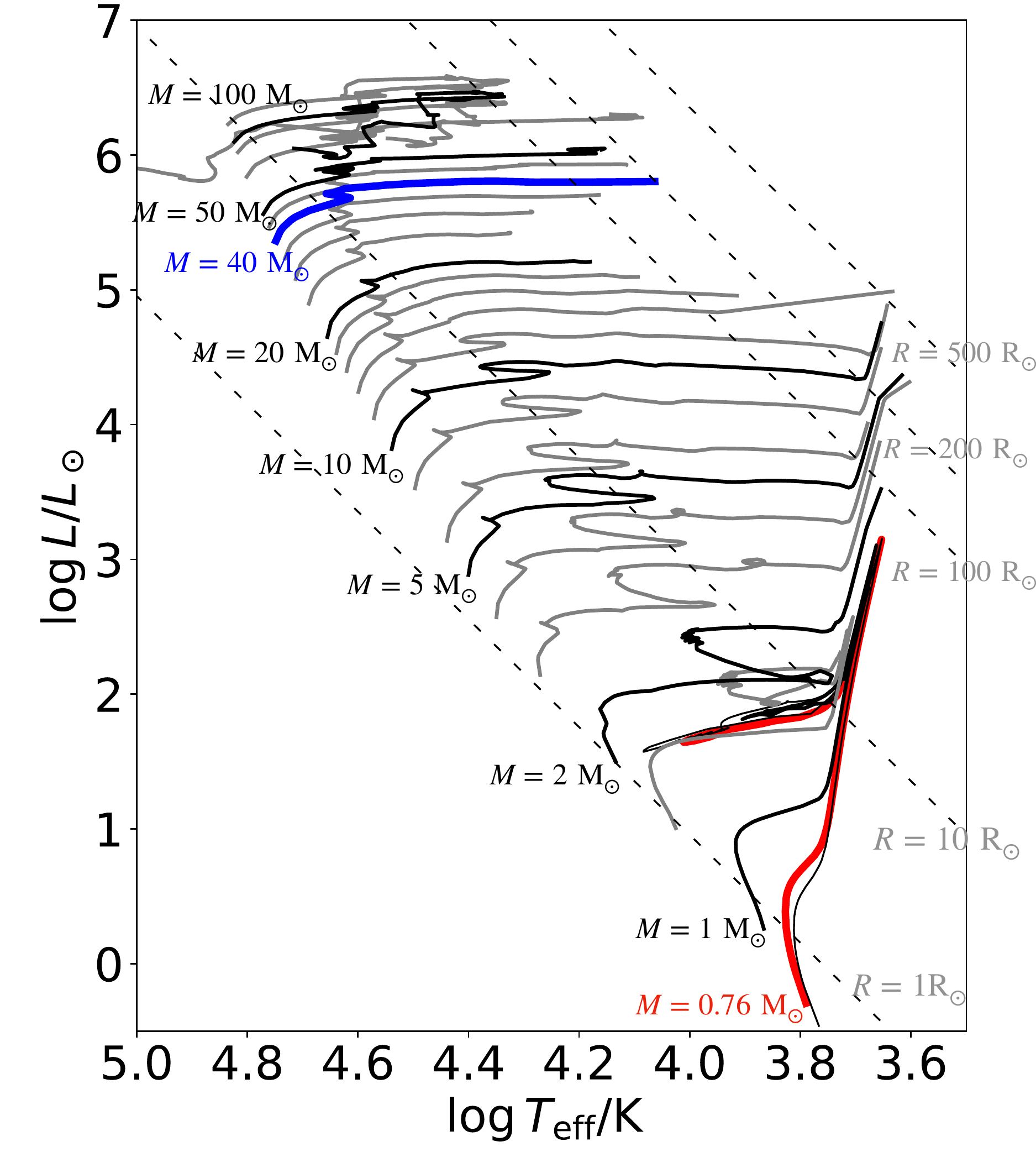}
      \caption{ Hertzsprung–Russell diagram of a sample of stars at metallicity $Z=4.1 \times 10^{-5}$  interpolated by {\sc sevn} from the {\sc mist} stellar tracks. The blue and red lines respectively show the likely progenitors of the \ac{BH} and stellar companion in {\it Gaia}~BH3.}
      \label{fig:misttrack}
   \end{figure}

In our fiducial model, we used the default {\sc sevn} options as described by \cite{iorio2023} (see also \citealt{sgalletta2023,costa2023}) except for the circularisation option. Here, we used the new default circularisation option \emph{angmom}, in which at the onset of the Roche-lobe overflow the orbit is circularised, conserving the angular momentum so that the new semi-major axis is $a_\mathrm{new}=a_\mathrm{old}(1-e^2_\mathrm{old})$. 

The stability of the Roche-lobe mass transfer is checked by comparing the current binary mass ratio to a critical mass ratio that depends on the stellar evolutionary phase. In this work, we used the  
default option of {\sc sevn} \citep{iorio2023} in which the Roche-lobe overflow episodes involving stars with a radiative envelope  (main-sequence stars and low-mass stars before climbing the red giant branch) are always stable, while in all the other cases we used the value of the critical mass ratio by \cite{hurley2002}. 
When the mass transfer is unstable, a common envelope evolution phase begins \citep{ivanova2013}. 
We used the standard $\alpha-\lambda$ prescription for the common envelope (\citealt{webbink1984,hurley2002}, and references therein) with the ejection efficiency parameter $\alpha=1$, whereas we estimated $\lambda$ using the formalism by \cite{claeys2014} and as implemented in {\sc sevn} (see Appendix A1.4 in \citealt{iorio2023}).

After a core-collapse supernova, we decided the mass of the BH based on the delayed model by \cite{fryer2012}. 
We adopted the supernova kick model by \cite{giacobbomapelli2020}, where the natal kick is drawn from a Maxwellian distribution with a one-dimensional root-mean-square $\sigma_\mathrm{rms}= 265 \ \mathrm{km/s}$ \citep{hobbs2005}, re-scaled by the ejecta mass  and the compact-remnant mass and resulting in a much lower  ($\lessapprox$ 10 km/s) effective kick for the massive \ac{BH}s ($>30$ M$_\odot$). 

In the fiducial model, we set the metallicity  to $Z=4.1 \times 10^{-5}$, which corresponds to the observed value of $[\textrm{Fe}/\textrm{H}]=-2.56$ when considering the {\sc mist} stellar models. 
The \textsc{mist} stellar tracks with the fiducial metallicity are shown in Fig. \ref{fig:misttrack}.

\subsection{Additional models: Parameter-space exploration} \label{sec:parexp}

To explore the parameter space for the formation of {\it Gaia}~BH3-like systems, we ran an additional set of \modref{17} simulations varying 
one parameter of the fiducial simulation per time \modref{ and two additional sets varying the initial conditions}. For the initial conditions, we used the same set of $2\times{}10^7$ binary systems as for the fiducial model. 

We simulated eight additional  metallicities: $Z=0.0001$ (model Z1E-4), $Z=0.0004$ (model Z4E-4), $Z=0.0008$ (model Z8E-4), $Z=0.001$ (model Z1E-3), $Z=0.004$ model (Z4E-3), $Z=0.006$ (model Z6E-3), $Z=0.008$ (model Z8E-3), and $Z=0.01$ (model Z1E-2).
We also explored the effect of the supernova kick by running four different simulations in which the kicks are drawn from a Maxwellian distribution with $\sigma_\mathrm{rms}= 1 \ \mathrm{km/s}$ (essentially no kick, model VK1), $50 \ \mathrm{km/s}$ (VK50), $100 \ \mathrm{km/s}$ (VK100), and $265 \ \mathrm{km/s}$ (VK265). In these models, the kick is not re-scaled for the ejected mass and the mass of the compact remnant. We also tested the importance of the common-envelope phase 
\modref{by running two simulations varying the common envelope efficiency ($\alpha=3$, model CE$\alpha$3, and $\alpha=5$, model CE$\alpha$5) and two additional simulations varying the binding energy formalism (model CE$\lambda$XL10 using the fitting equations by \citealt{xuli2010} and model CE$\lambda$K21 using the fitting equations by \citealt{klencki2021}\footnote{\modref{In {\sc sevn} version 2.10.0, we revised the implementation of the $\lambda_\mathrm{K21}$ fitting equations by \citealt{klencki2021}. We corrected a bug in the code that affected the $\lambda_\mathrm{K21}$ estimate for  metallicities $Z<0.017$ (the bug has been fixed for all the versions greater than 2.8.0). Additionally, we improved the fit for cases in which the stellar radius is less than the radius at the terminal age main sequence (TAMS) of the {\sc mesa} stellar tracks used in \cite{klencki2021} (private communication). In this radial range, the original fitting equations were extrapolated, leading to high binding energies ($> 10^{52}$ ergs). For additional information, we refer to the discussion in the {\sc sevn} repository: \url{https://gitlab.com/sevncodes/sevn/-/issues/4}. }}).}  
\modref{Furthermore,} we ran an additional model (RLOp) in which we allowed the system to start a Roche-lobe overflow in an eccentric orbit if the condition of Roche-lobe overflow is satisfied at the periastron, that is, if the stellar radius is larger than 
$R_\mathrm{L}=r_\mathrm{peri}f(q)$, where $r_\mathrm{peri}$ is the periastron and $f(q)$ is a function that depends on the binary mass ratio \citep{eggleton1983}.

\modref{Since {\it Gaia}~BH3 is an eccentric system, we also ran an additional simulation (model ICthermal) to examine the impact of the initial eccentricity distribution. In this simulation, the initial orbital eccentricity is drawn from a thermal distribution, which favours eccentric systems, PDF$(e) \propto e$ \citep[see e.g.][]{geller2019}.}

Finally, to assess the uncertainties due to the initial condition sampling, we drew an additional set of \modlref{$1.8\times{}10^8$} binaries and simulated their evolution using the same setup as the fiducial simulation. \modlref{We combined the results of the additional simulations with those from the fiducial simulation, and labeled the final combined dataset as `fiducial2E8'.} 
The total mass evolved in this case is $M_\mathrm{sim}=4.2\times10^9 \ \mathrm{M}_\odot$, corresponding to a total population mass of $M_\mathrm{tot}=1.4\times10^{10} \ \mathrm{M}_\odot$ (see Sect. \ref{sec:fiducial}).


\section{Results}\label{sec:results}
   \begin{figure*}
   \centering
   \includegraphics[width=\hsize]{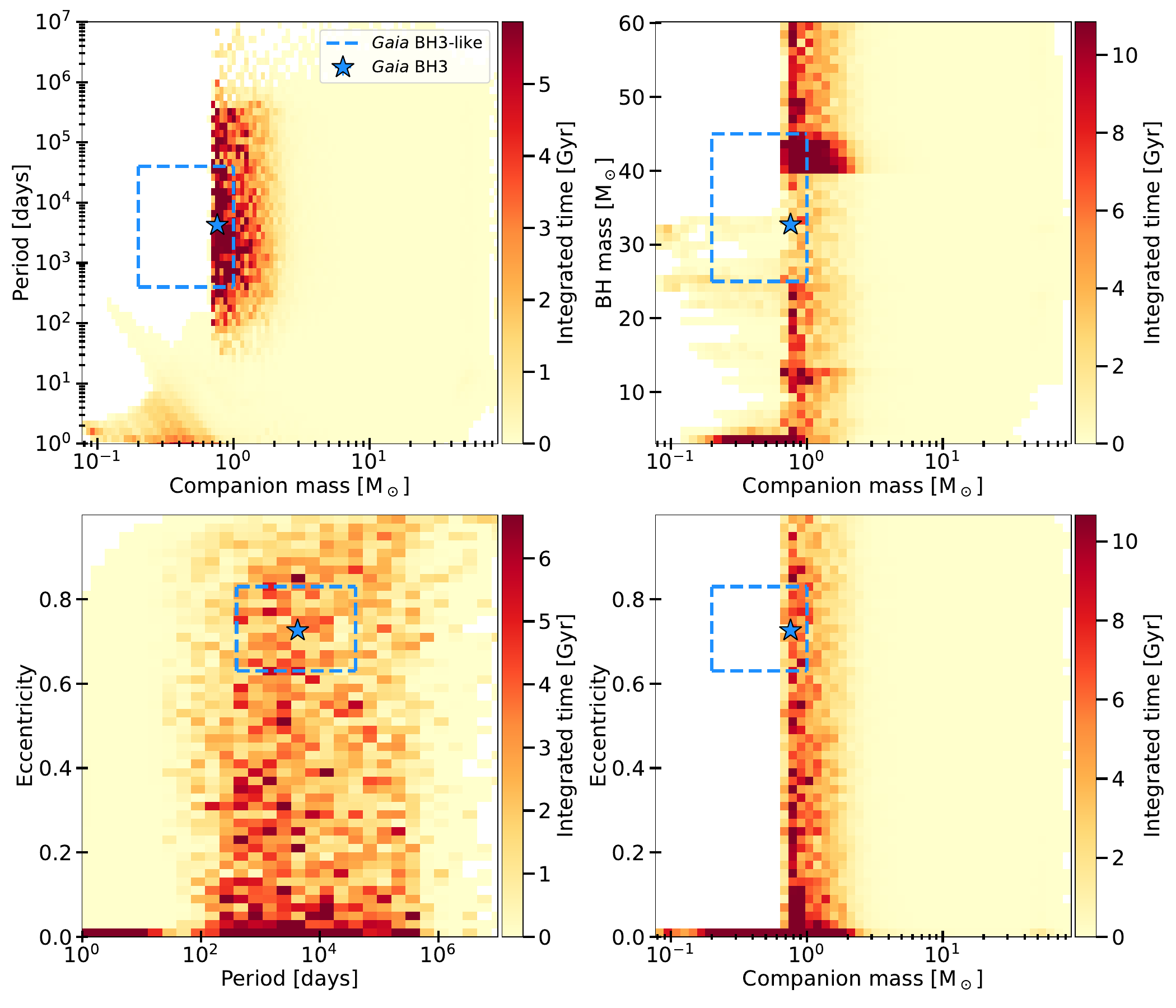}
      \caption{Properties of non-interacting BH-star systems in the fiducial model. Here, we only show stars in the giant phase, that is, outside the main sequence and that have not started the core helium burning phase yet. Also, we do not show systems that undergo a Roche-lobe overflow episode. From top to bottom and from left to right: Orbital period versus companion star mass; \ac{BH} mass versus companion star mass; eccentricity versus orbital period; eccentricity versus companion star mass. The colour map shows the integrated time spent by the systems in a given bin; thus, it is a proxy of the likelihood of observing a system with such properties. The map does not include information about the stellar luminosity, which is also important for detectability. We assumed a metallicity of $Z=4.1 \times 10^{-5}$. 
      The star indicates the properties of {\it Gaia}~BH3, while the light blue boxes show the criteria used to select {\it Gaia}~BH3-like systems. The colour maps saturate at 98\% of the maximum value. 
      }
         \label{fig:orb}
   \end{figure*}

Figure~\ref{fig:orb} shows the properties of  the BH-star systems in the fiducial model. 
We considered all the simulation outputs in which a binary system is composed of a BH and a giant star, that is, a star that left the main sequence and has not started the core helium burning phase yet. We imposed the additional condition that the system is not undergoing a common-envelope or Roche-lobe overflow episode. 
The colour map of Fig.~\ref{fig:orb} 
represents the time spent by all the systems in such a configuration, which is a proxy of the likelihood of observing a given system. 

Binaries hosting a BH and a massive companion are, in general, more common in the simulations, but the short lifetime of the massive star makes them less likely to be observed. 
In contrast, BH--low-mass star binaries are rarer, but once formed, they can remain in the same configuration for several \modlref{gigayears}. 

{\it Gaia}~BH3 is located in a region of the parameter space well covered by the fiducial simulation, especially when considering binary systems older than 10 Gyr. 
In addition, the {\it Gaia}~BH3 properties (mostly period and BH mass) cover a range in which {\it Gaia} is particularly sensitive to astrometric signals of binarity  \citep[see e.g.][]{penoyre2022}.
From Fig. \ref{fig:orb}, we can conclude that if a BH-star system is observed in an old and metal-poor population, it is not surprising to find it with the {\it Gaia}~BH3 configuration. In this sense, {\it Gaia}~BH3 is not an outlier,
unlike {\it Gaia} BH1 and BH2 \citep{elbadry2023a,rastello2023}. 

It is convenient to define a region of the observed parameter space to identify {\it Gaia}~BH3-like systems. 
For the sake of simplicity and to help the comparison with previous work, we used a selection cut \modref{similar to the one} used in \cite{marinpina2024}: $P \in [400,40000] \ \mathrm{days}$, $m_{s} \in [0.2,1] \ \mathrm{M}_\odot$, $m_\mathrm{bh} \in [25,45] \ \mathrm{M}_\odot$, $e \in [0.63,0.83]$. The box represents about 1\% of the volume of the parameter space that includes BH-star systems at ages older than 10 Gyr. 

\subsection{Formation history} \label{sec:form_hist}

All the {\it Gaia}~BH3-like systems in our models have the same simple formation history. Their progenitors consist of a massive star ($>40$ M$_\odot$) and a low-mass companion ($<1$ M$_\odot$). The binary components never interact through Roche-lobe overflow, so the orbital properties at BH formation are almost the same as their initial ones. The final properties of the BH-star system are thus determined by the binary phase at the moment of the supernova explosion, the amount of mass lost during the evolution, and the BH natal kick.

In our fiducial model, the initial BH progenitor mass is about 40--50 M$_\odot$, and the ejected mass at BH formation is 10--20 M$_\odot$. The mild BH natal kick ($<10$ km/s) cannot drastically change the binary orbital parameters. As a consequence, a large majority of BH progenitors originated with orbital properties very similar to {\it Gaia}~BH3, namely, a period between $10^3$ and $10^4$ days and an eccentricity in the range $0.6-0.9$. 
The system more similar to {\it Gaia}~BH3 has an initial period of about 1500 days, an eccentricity of 0.65, a stellar mass of 0.83 M$_\odot$, and an initial BH progenitor mass of 41 M$_\odot$.
We found a sub-population ($<10$\% of the total) that produces {\it Gaia}~BH3-like systems starting from almost circular orbits ($e<0.2$) and lower initial periods ($<10^3$ days). For these systems, the final orbital properties are determined by the high value of the BH natal kick. A few systems ($\approx{}3\%$) have very massive BH progenitors of about 120 M$_\odot$ that produce BHs 
with a mass of $\sim{30}$~M$_\odot$ because of  mass loss during pulsational pair instability \citep{heger2002,belczynski2016,woosley2017,woosley2019,farmer2019,costa2021,tanikawa2021b}.

\modref{In all of our models, the  progenitors of {\it Gaia}~BH3 do not 
undergo any (stable or unstable) mass transfer episode. Therefore, there are no differences among the models CE$\alpha3$, CE$\alpha5$, CE$\lambda$XL10, and CE$\lambda$K21}. 
\modref{Also, we found only a mild decrease in the number of {\it Gaia}~BH3-like systems  for the model RLOp}, which indicates that including or not including the possibility to trigger a stable mass transfer at periastron is not important for the production of such binary systems.
We found the same progenitors in the case of the kick model VK1, which serves as confirmation that in the fiducial model, the implemented kick model produces very low kicks for the case of {\it Gaia}~BH3-like progenitors.

Models with higher kick velocities show significant differences in the initial orbital properties. The BH natal kicks in models VK50 and VK100 are able to wash out any correlation between the initial and final period and eccentricity.
The {\it Gaia}~BH3-like progenitors have initial periods and eccentricities almost uniformly distributed in the range 100--4000 days and 0--0.8, respectively.
At very high velocities (model VK265), the formation of {\it Gaia}~BH3-like binaries is strongly suppressed. Most systems break during BH formation, and just a few of them (four out of the 20 million simulated binaries) had a lucky configuration of binary phase and/or velocity kick that let them survive the supernova event.

We found a trend with metallicity for the initial orbital period, eccentricity, and BH progenitor mass. From low metallicity (model Z1E-4) to intermediate-high metallicity (model Z1E-2), the lower limit of the initial period moves toward higher values ($P>1000$  days for model Z1E-2), while the sub-populations with initial low eccentricity ($e<0.6$) tend to disappear. As for the BH progenitor mass, the range of values widens with increasing metallicity, up to initial masses of 100 M$_\odot$. This is due to the non-monotonic trend of the initial BH mass relation of the {\sc mist} tracks used in this work when paired with the rapid and delayed supernova models by \cite{fryer2012}.
When assuming initial masses lower than 60 M$_\odot$, the combination of the \cite{fryer2012} supernova models and the {\sc mist} tracks produces BHs more massive than 25 and 30 M$_\odot$ up to a metallicity $Z=0.02$ and $0.01$, respectively. When relaxing the limit on the initial masses, BHs more massive than 30 M$_\odot$ are excluded only for $Z>0.02$. 

\modref{
   Mass transfer through stellar winds is negligible ($<10^{-9}$ M$_\odot$) in all of the simulated {\it Gaia}~BH3-like objects.  
At BH formation, a typical {\it Gaia}~BH3 progenitor hosts a low-mass star with stellar radius $R_*<1$ R$_\odot$ at  a distance from the supernova explosion of $D_\mathrm{sn}=1000$ -- $4000$ R$_\odot$. Considering a total ejected mass of $M_\mathrm{ej}=10$ M$_\odot$, the amount of mass deposited into the low-mass companion is $<M_\mathrm{ej} R^2_*/D^2_\mathrm{sn}\sim$ $10^{-5}$--$10^{-6}$ M$_\odot$.  As discussed by \cite{elbadry2024}, this amount of pollution is hardly detectable in the spectral lines.
In addition, the {\sc MIST} models predict that a $\sim 0.8$ M$_\odot$ star loses $10^{-2}$--$10^{-3}$ M$_\odot$ from the main-sequence to the red giant phase, likely removing all the material deposited from the supernova into the star external layer. In conclusion, we do not expect any peculiar chemical signature in the {\it Gaia}~BH3 star if it formed through the isolated channel. This is consistent with the observations \citep{GaiaBH3}.
}

   \begin{figure*}
   \centering
   \includegraphics[width=0.9\hsize]{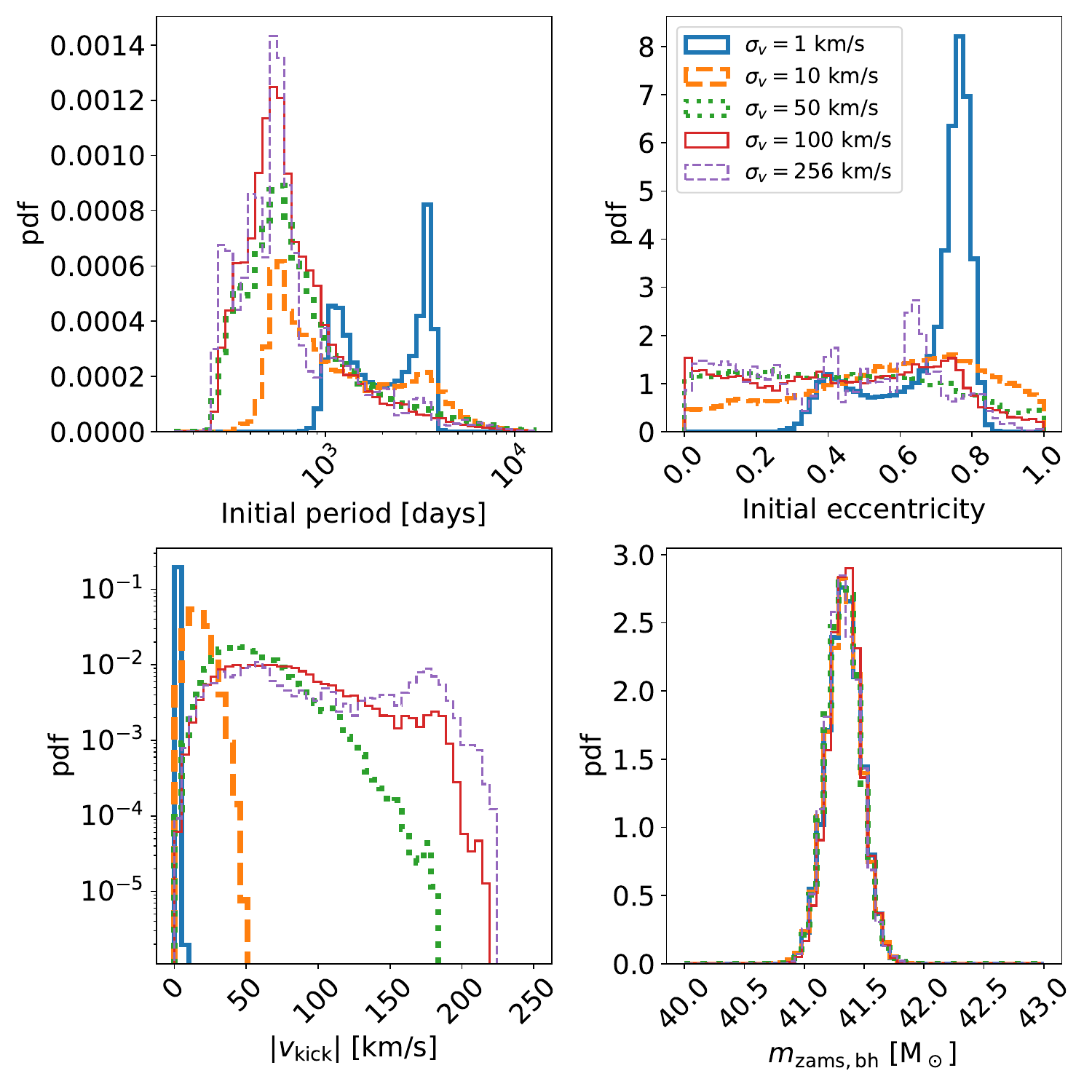}
      \caption{Distribution of the initial conditions of the binaries that reproduce the {\it Gaia}~BH3 properties after  BH formation. From top left to bottom right, the panels show the initial binary period, initial eccentricity, the module of the BH natal kick velocity, and the initial mass of the BH progenitor. We show the distributions obtained assuming different Maxwellian-distribution priors for the kick velocity: $\sigma_v=1 \ \mathrm{km/s}$ (blue thick solid line), $\sigma_v=10 \ \mathrm{km/s}$ (orange thick dashed line), $\sigma_v=50 \ \mathrm{km/s}$ (green dotted line), $\sigma_v=100 \ \mathrm{km/s}$ (red thin solid line), $\sigma_v=256 \ \mathrm{km/s}$ (violet thin dashed line).}\label{fig:icsample}
   \end{figure*}

\subsection{Bayesian sampling}
The \textsc{sevn} simulations indicate that the region of {\it Gaia}~BH3 in the observed parameter space ($P$, $e$, $m_\mathrm{BH}$, $m_\mathrm{ms}$) tends to be populated by systems in which the binary components never interact. Therefore, the final binary properties are determined by the initial binary parameters and the details of the BH formation only.
In order to focus on  {\it Gaia}~BH3 progenitors independently of the chosen selection cut, 
\modlref{we implemented a simple Bayesian sampling technique in which we sample the posteriors of the binary initial conditions. We assume the likelihood is a multivariate normal distribution centred on the {\it Gaia}~BH3 values and with the standard deviations equal to the reported errors (which we assumed to be independent).}

For the parameter \modlref{priors}, we used:
\modlref{
\begin{itemize}
\item a uniform distribution for the mean anomaly (between $0$ and $2 \pi$) at the moment of the BH formation,
\item a uniform distribution for the mass of the BH progenitor between 20 and 100 M$_\odot$,
\item a uniform distribution for eccentricity between 0 and 1,
\item a log-uniform distribution for the period between 10 and $10^6$ days,
\item a normal distribution for the three spherical components of the velocity kick, corresponding to a Maxwellian distribution of the velocity module.
\end{itemize}
}
\modlref{For the velocity distribution,} we tested five different  values of the one-dimensional root-mean-square velocity for the Maxwellian distribution ($1, 10, 50, 100, 256$ km/s). We also excluded the interacting binary systems by setting the prior to be equal to zero when the maximum radius of the \ac{BH} progenitor is larger than either the separation at periastron or the Roche-lobe radius at periastron (see Sect. \ref{sec:method}). To estimate the final properties of the binary, we used the formalism in the appendix of \cite{hurley2002}, and we estimated the maximum radius and the final mass of the \ac{BH} progenitor by using {\sc sevn}.
The posteriors were sampled using the Monte Carlo Markov chain (MCMC) ensemble sampler available in the `emcee' python package \citep{foreman2013}. This method is similar to the grid search presented by \cite{elbadry2024}, but our method is tailored to exactly reproducing the observed properties of {\it Gaia}~BH3. Since we included the reduction of the parameter space due to  stellar interactions directly in the priors, the posteriors do not need to be corrected a posteriori.

We show the posterior samples in Fig. \ref{fig:icsample}. Overall, they are consistent with what was obtained from the {\sc sevn} simulations for the wider space of the selection cut assumed in Sect. \ref{sec:form_hist}. Assuming low kick velocities, the {\it Gaia}~BH3 progenitor is very likely to have originated from a wide and eccentric binary ($P\approx 3500$ days, $e\sim0.75$). The smaller peak at around $10^3$ days and $e \sim 0.4$ is due to systems that form the \ac{BH} at periastron, where it is more favourable to produce large final eccentricities due to the impact of mass loss during the supernova explosion. Larger kicks widen the range of initial periods and eccentricities, with most of the initial periods being around 500--600 days. The period distribution is sharply cut below 300 days because the objects are tight enough to interact. It is possible to form a {\it Gaia}~BH3 system with initial kicks up to 200 km/s. For higher values, the kick is larger than the escape velocity at periastron in the whole range of periods \modlref{that are} consistent with a non-interacting binary.

Independent of the explored parameters, a clear picture emerged. In the presence of low supernova kicks ($< 10$ km/s), {\it Gaia}~BH3-like systems form from non-interacting binary progenitors preferentially in a wide ($500 <P/\mathrm{days}<2\times{}10^4$) and eccentric configuration. 
Alternatively, on the off chance that the binary survives to higher supernova kicks, systems with {\it Gaia}~BH3 properties can form preferentially from tighter ($P/\mathrm{days}\sim 500$) systems almost independently of their initial eccentricity if $e<0.7$.
For this reason, the initial properties of the binary can play a key role in setting the formation efficiency of {\it Gaia}~BH3-like systems. \modref{Indeed, the model with a higher initial eccentricity (ICthermal) produces two to three times more {\it Gaia}~BH3-like systems compared to our fiducial run (Table \ref{tab:feff}).}

\subsection{Halo population} \label{sec:halopop}

\begin{table*}[]
\caption{Formation efficiency (M$^{-1}_\odot$) for {\it Gaia} BH3-like and BH-star systems obtained from different simulation models.}
\centering
\begin{tabular}{cccccc}
\hline
Model    & $\eta_\mathrm{{\it Gaia}~BH3-like}$ & \begin{tabular}[c]{@{}c@{}}$ \eta_\mathrm{{\it Gaia}~BH3-like}$\\ ($t>10$  Gyr)\end{tabular}  &  \begin{tabular}[c]{@{}c@{}}$ \eta_\mathrm{{\it Gaia}~BH3-like}$\\ ($t>10$   Gyr) \\ ($v_\mathrm{sys}<10$ km s$^{-1}$) \end{tabular} &  \begin{tabular}[c]{@{}c@{}}$ \eta_\mathrm{{\it Gaia}~BH3-like}$\\ ($t>10$   Gyr) \\ ($v_\mathrm{sys}<1$ km s$^{-1}$) \end{tabular} &  \begin{tabular}[c]{@{}c@{}}$\eta_\mathrm{BH-star}$\\ ($t>10$   Gyr)\end{tabular} \\ \hline
fiducial & $(9.7\pm0.8) \times  10^{-8}$                   & $(3.9\pm0.5) \times  10^{-8}$                &                                             $(2.8\pm0.4) \times  10^{-8}$     &        $(2.5\pm0.4) \times  10^{-8}$                          & $(1.6\pm0.1) \times  10^{-6}$                                                                             \\
\modref{fiducial2E8} & $(1.0\pm0.1) \times  10^{-7}$                   & $(4.4\pm0.2) \times  10^{-8}$                &                                             $(3.5\pm0.2) \times  10^{-8}$                 &      $(3.0\pm0.2) \times  10^{-8}$                   & $(1.6\pm0.1) \times  10^{-6}$         \\
\modref{ICthermal} & $(2.4\pm0.1) \times  10^{-7}$                   & $(1.0\pm0.1) \times  10^{-7}$                &                                             $(9.0\pm0.8) \times  10^{-8}$                       &   $(7.8\pm0.7) \times  10^{-8}$               & $(1.3\pm0.1) \times  10^{-6}$                                                                        \\
RLOp     &   $(9.1\pm0.8) \times  10^{-8}$                   & $(3.5\pm0.5) \times  10^{-8}$        &                                                     $(2.5\pm0.4) \times  10^{-8}$                &    $(2.3\pm0.4) \times  10^{-8}$                       & $(1.6\pm0.1) \times  10^{-6}$                                                                         \\ \hline
VK1      &  $(1.0\pm0.1) \times  10^{-7}$                        & $(3.9\pm0.5) \times  10^{-8}$       &                         $(3.9\pm0.5 )\times  10^{-8}$        &  $(8.5\pm0.2) \times  10^{-9}$   & $(2.3\pm0.1) \times  10^{-6}$          \\
VK50     &  $(9.0\pm0.8) \times  10^{-8}$                     & $(4.0\pm0.5) \times  10^{-8}$                 &                                                      0                 & 0      & $(9.0\pm0.3) \times  10^{-7}$                                 \\
VK100    & $(3.6\pm0.5) \times  10^{-8}$                     & $(1.1\pm0.3) \times  10^{-8}$          &                                                             0                    & 0        & $(6.0\pm0.2) \times  10^{-7}$                                                                             \\
VK265    &    $(6.4\pm2.1) \times  10^{-9}$            &                                              $(2.1\pm1.2) \times  10^{-9}$       &                   0               & 0   &  $(6.2\pm0.2) \times  10^{-8}$  \\      \hline                                     
CE$\alpha$3     &     $(9.7\pm0.8) \times  10^{-8}$              &          $(3.9\pm0.5) \times  10^{-8}$  &    $(2.8\pm0.4) \times  10^{-8}$        &   $(2.5\pm0.4) \times  10^{-8}$ &$(2.5\pm0.1) \times  10^{-6}$   \\ 
CE$\alpha$5     &        $(9.7\pm0.8) \times  10^{-8}$            &     $(3.9\pm0.5) \times  10^{-8}$      &    $(2.8\pm0.4) \times  10^{-8}$    &  $(2.5\pm0.4) \times  10^{-8}$     &   $(2.8\pm0.1) \times  10^{-6}$    \\
\modref{CE$\lambda$XL10}     &        $(9.7\pm0.8) \times  10^{-8}$            &     $(3.9\pm0.5) \times  10^{-8}$      &    $(2.8\pm0.4) \times  10^{-8}$    &  $(2.5\pm0.4) \times  10^{-8}$     &   $(1.4\pm0.1) \times  10^{-6}$  \\
\modref{CE$\lambda$K21}    &        $(9.7\pm0.8) \times  10^{-8}$            &     $(3.9\pm0.5) \times  10^{-8}$      &    $(2.8\pm0.4) \times  10^{-8}$    &  $(2.5\pm0.4) \times  10^{-8}$      &   $(1.4\pm0.1) \times  10^{-6}$   \\
\hline
Z1E-4    &          $(1.3\pm0.1) \times  10^{-7}$            &       $(4.8\pm0.6) \times  10^{-8}$                                              &               $(3.0\pm0.5) \times  10^{-8}$                &    $(2.7\pm0.4) \times  10^{-8}$             &   $(1.4\pm0.1) \times  10^{-6}$   \\
Z4E-4    &       $(1.1\pm0.1) \times  10^{-7}$             &                $(4.8\pm0.6) \times  10^{-8}$                                         &          $(2.1\pm0.4) \times  10^{-8}$               &    $(1.7\pm0.3) \times  10^{-8}$             &  $(1.4\pm0.1) \times  10^{-6}$  \\
Z8E-4     &     $(9.4\pm0.8) \times  10^{-8}$                &                                              $(3.9\pm0.5) \times  10^{-8}$         &                $(1.8\pm0.4) \times  10^{-8}$      &     $(1.4\pm0.3) \times  10^{-8}$                    &  $(1.5\pm0.1) \times  10^{-6}$    \\
Z1E-3     &    $(9.2\pm0.8) \times  10^{-8}$               &                                                $(4.0\pm0.5) \times  10^{-8}$                       &      $(2.1\pm0.4) \times  10^{-8}$     &        $(1.9\pm0.4) \times  10^{-8}$                          &  $(1.6\pm0.1) \times  10^{-6}$  \\
Z4E-3     &    $(9.1\pm0.8) \times  10^{-8}$                &                                         $(5.3\pm0.6) \times  10^{-8}$                       &             $(2.8\pm0.5) \times  10^{-8}$         &       $(2.1\pm0.4) \times  10^{-8}$           &  $(2.4\pm0.1) \times  10^{-6}$   \\ 
Z6E-3     &     $(9.4\pm0.8) \times  10^{-8}$                   &                                  $(5.8\pm0.6) \times  10^{-8}$                       &          $(3.9\pm0.5) \times  10^{-8}$                     &     $(2.8\pm0.4) \times  10^{-8}$         & $(2.7\pm0.1) \times  10^{-6}$     \\ 
Z8E-3     &      $(8.2\pm0.8) \times  10^{-8}$               &                                                $(6.2\pm0.7) \times  10^{-8}$               &        $(4.7\pm0.6) \times  10^{-8}$                 &    $(3.3\pm0.5) \times  10^{-8}$          &  $(3.1\pm0.1) \times  10^{-6}$  \\
Z1E-2     &      $(8.2\pm0.8) \times  10^{-8} $            &                                                    $(7.1\pm0.7) \times  10^{-8}$       &    $(7.0\pm0.7) \times  10^{-8}$       &  $(5.9\pm0.6) \times  10^{-8}$     &  $(3.4\pm0.1) \times  10^{-6}$    \\ \hline
\end{tabular}
\tablefoot{Column~1: Model name; Column 2: Formation efficiency of all {\it Gaia}~BH3-like systems, as defined by the section cut (Sect.~\ref{sec:results}); Col. 3: Same as Col.~2 but with the additional condition that the system is still in the BH-star configuration after 10 Gyr; Col. 4 (5): Same as Col.~3 but with the additional condition that the binary has a systemic velocity of $v_\mathrm{sys}<10$ km/s ($v_\mathrm{sys}<1$ km/s); Col. 6: All BH-star systems with an age $>10$ Gyr. To calculate the formation efficiency, we assumed a 100\% binary fraction. The errors are estimated as the square root of the counts over the total mass.}
\label{tab:feff}
\end{table*}

In the explored models,  {\it Gaia}~BH3-like systems represent one of the standard outcomes of binary evolution. 
Next, we investigated how common {\it Gaia}~BH3-like systems and BH-star binaries are in the Galactic halo environment, which is characterised by low metallicity and old ages. We also assumed  $t>10$ Gyr.
First, we estimated the formation efficiency, defined as the ratio of the number of objects to the total simulated mass.
We corrected the total mass in the initial conditions for the incomplete mass sampling of the primary, and  we assumed an initial binary fraction of one for the sake of simplicity. 
Table \ref{tab:feff} reports the formation efficiency of all the systems in the {\it Gaia}~BH3 selection box (Fig. \ref{fig:orb}) as well as of those systems that are in the selection box and in the BH-star configuration after 10 Gyr.

In the fiducial model, the population of {\it Gaia}~BH3-like objects accounts for about 2\%-3\% of the whole population at $t>10$ Gyr. This number is consistent with the volume fraction sampled by the selection box. Models with higher kicks significantly suppress the formation of both {\it Gaia}~BH3-like and BH-star systems, while a higher metallicity increases their formation efficiency up to a factor of two. 
Given the formation history of {\it Gaia}~BH3-like objects, the efficiency is not affected 
by the choice of the common-envelope ejection efficiency \modref{and binding energy}. However, 
larger values of the common envelope efficiency parameter $\alpha$ slightly increase the formation efficiency of the whole BH-star systems, resulting in a lower fraction of {\it Gaia}~BH3-like systems.

Even if the Galactic stellar halo is the least massive component  of our Galaxy, the most recent estimates revise its mass from a few $10^8$ M$_\odot$ (see, e.g., \citealt{bell2008})  up to  $0.7-1.2 \ \times 10^9$ M$_\odot$ \citep{deason2019,lane2023}.
Combining the total halo mass with the formation efficiencies 
in Table \ref{tab:feff}, we predict the number of \modlref{BH stars} in the halo to be in the range of  $\approx 1000-4000$. \modref{Among these systems, about $20-120$} fall in the {\it Gaia}~BH3 selection cut used in this work (Fig. \ref{fig:orb}). 
The Galactic stellar halo is a diffuse component extended up to 100 kpc with a steep density profile ($\rho \propto r^{-\gamma}$, with $2<\gamma<5$; see, e.g., \citealt{medina2024,iorio2018} and references therein).

\cite{deason2019} report a local density profile at the Sun location of  $ \rho_\mathrm{h,\odot} \approx 8 \times 10^{4} \ \mathrm{M}_\odot/\mathrm{kpc}^3$.
Multiplying this density by the formation efficiencies of the old BH-star binaries, we obtained a local density of 
$n_\mathrm{BHStar}=\eta_\mathrm{BHStar} \, \rho_\mathrm{h,\odot}$ ranging from 0.1  to 0.3 $\mathrm{kpc}^{-3}$. The values correspond to 
an expected number of objects between 0.13 and 0.27 at the distance of {\it Gaia} 
BH3 ($\approx 0.6  \ \mathrm{kpc}$). At face value, we do not expect any BH-star binary system from the halo in the solar neighbourhood. However, we can account for the statistical noise by assuming that the number of systems follows a Poissonian distribution with the rate $\lambda_\mathrm{BHStar}$ ranging between 0.13 and 0.27. In this case, the probability to observe at least one object is between 12\% and 25\%. 
By simply using the formation efficiency for the old systems in the {\it Gaia}~BH3 selection box (Fig. \ref{fig:orb}), we obtained a Poisson rate between 0.003 and 0.006, corresponding to a probability of observing at least one {\it Gaia}~BH3-like system of 0.3--0.6\%. Although these numbers correspond to the very low probability tail of the distribution, they cannot be rejected at the 3$\sigma$ level. However, a single future detection of another {\it Gaia}~BH3-like binary system in the Galactic halo within the solar neighbourhood would be enough to significantly rule out the isolated formation channel, according to the models presented in this work. 
However, this will also represent a challenge for the alternative dynamical  formation channel (see Sect. \ref{sec:dyn}).

\modref{

\subsection{Populations in the stellar clusters and cold structures} \label{sec:cluster}

\cite{balbinot2024} show that {\it Gaia}~BH3 is associated with  the ED-2 stellar stream.
The presence of {\it Gaia}~BH3-like systems in star clusters and kinematically cold structures, such as a stellar stream, pose additional constraints on the natal kick velocity in the binary centre of mass, $v_\mathrm{sys}$. In fact, if the kick velocity is significantly larger than the typical velocity dispersion of the host cluster or stream, the BH will escape from the system and join the field population in the Milky Way. The escape velocity from globular clusters ranges from  $\sim10 \ \mathrm{km \ s}^{-1}$ km to $\sim50\ \mathrm{km \ s}^{-1}$ \citep{gnedin2002}. Almost all ($>90 \%$) the {\it Gaia} BH-3 systems in our simulations have $v_\mathrm{sys}<50\ \mathrm{km \ s}^{-1}$, except for models VK50, VK100, and VK265 (Fig. \ref{fig:icsample}).

Focusing on the low-velocity tail of the kick distribution, Table~\ref{tab:feff} reports the formation efficiency of {\it Gaia}~BH3-like systems when we apply the additional selection cut of $v_\mathrm{sys}~<~10~\mathrm{km \ s}^{-1}$ (typical velocity dispersion of globular clusters) and $v_\mathrm{sys}~<~1~\mathrm{km \ s}^{-1}$ (drift velocity of the ED-2 components with respect to the progenitor cluster;  \citealt{marinpina2024}). For most of our models, the formation efficiency decreases by $\leq{}20-30$\%. Although the effect is not negligible, 
the  formation efficiency decrease is not significant enough to rule out the isolated formation channel for {\it Gaia}~BH3-like systems at a 3$\sigma$ level.

}

\section{Discussion}\label{sec:discussion}

\subsection{Stellar evolution models} \label{sec:elbadry}

\modref{
Uncertainties about massive star evolution  have an important impact on our results. For instance, differences in the maximum radius of the BH progenitor are crucial to selecting which objects interact during binary evolution.

\cite{elbadry2024} explored the main uncertainties about stellar evolution models, 
concluding that most models predict that the progenitor of {\it Gaia}~BH3 reached radii up to 1000~R$_\odot$. Such stars cannot avoid interaction before BH formation. For this reason, \cite{elbadry2024} conclude that dynamical interactions in a star cluster represent the only possible formation channel for  {\it Gaia}~BH3, unless the radius of the BH progenitor did not exceed $\approx 700$  R$_\odot$. 
In our stellar evolution models based on the \textsc{MIST} tracks \citep{choi2016}, the BH progenitor reaches a maximum radius of $\lesssim{}200-300$ R$_\odot$ for $Z<0.0002$ (Fig.~\ref{fig:misttrack}). We find similar results for the {\sc parsec} stellar tracks \citep{bressan2012,chen2015,costa2021,nguyen2022} available in {\sc sevn} \citep{iorio2023}. While a detailed analysis of the differences in stellar evolution models is beyond the scope of this paper (see, e.g.,  \citealt{agrawal2020,agrawal2022b,agrawal2022,klencki2020,Gilkis2021}), it is still crucial to note their significant impact on the interpretation of {\it Gaia}~BH3 and other BH-star systems.
}

\subsection{Initial eccentricity and stellar tides}

In all the simulations assuming our fiducial kick model \citep{giacobbomapelli2020}, {\it Gaia}~BH3-like systems form preferentially from binary systems with an initial eccentric orbit ($e>0.6$). 
Therefore, we expect that initial distributions that favour eccentric orbits, such as the thermal distribution, boost the number of {\it Gaia}~BH3-like systems (model ICthermal; see Table~\ref{tab:feff}). In contrast, circular orbits significantly suppress the formation of such systems unless high kick models are used. 

In some cases, initial circular binaries are justified by the effect of tides. The {\sc sevn} simulations used in this work implement the tides following the formalism by \cite{hurley2002} (see also \citealt{iorio2023}) and show that for {\it Gaia}~BH3-like systems, tides are extremely inefficient. 
 The tidal efficiency is at its maximum after the main-sequence phase of the BH progenitor, when the star expands and develops a convective envelope. 
 Assuming that the dominant tidal dissipation mechanism is due to convective motions in the envelope \citep{zahn1977,hurley2002} and considering the {\sc mist} models at $Z=4.1 \times 10^{-5}$ (Fig. \ref{fig:misttrack}), 
 we estimated a circularisation timescale due to tides of $\tau_{\rm circ} \simeq 10^{25}$--$10^{38} \,\rm Gyr$.  This timescale is orders of magnitude longer than the lifetime of the binary system and especially longer than the post main-sequence phase, when tides are more efficient.

\subsection{Systematic uncertainties}

Table~\ref{tab:feff} shows that the formation efficiency of BH-star systems drops significantly for models with high supernova kicks. In the models VK100 and VK265, we can exclude the hypothesis that the observation of {\it Gaia}~BH3 is consistent with the isolated channel at a high level of significance ($>3 \sigma$). 
Estimates of BH kicks rely on a few tracers (e.g. X-ray binaries) and are subject to many uncertainties and systematics (see, e.g., \citealt{andrews2022,atri2019,shenar2022a}).
The association of {\it Gaia}~BH3 with a stellar stream \citep{balbinot2024} 
indicates that the binary system did not experience a kick much higher than the typical escape velocity from a star cluster ($< 10$ km/s) at the time of BH formation \modref{(Sect. \ref{sec:cluster})}.

In our analysis, we did not consider binaries with companion stars less massive than $0.71 \ \mathrm{M}_\odot$. As discussed by \cite{elbadry2024}, such a population can represent a significant portion of the stellar companions of {\it Gaia}~BH3-like systems. 
To assess the relevance of this population on the estimate of the formation efficiency, we generated new initial conditions, as done in Sect. \ref{sec:ic}, but we assumed an extended mass ratio range from $0.08\,{}\textrm{M}_\odot/m_1$ to 1.
We find that the percentage of binaries hosting a BH progenitor and a secondary star less massive than $0.71 \ \mathrm{M}_\odot$ is 40\% of similar binaries with a secondary in the mass range $0.71-1.0 \ \mathrm{M}_\odot$. 
Therefore, we expect an overall increase in the formation efficiency by about 40\% with respect to the value reported in Table \ref{tab:feff}. 
This factor is balanced by the commonly assumed binary fraction in stellar populations ($40-60\%$; see, e.g., \citealt{moe2017}).

Many population synthesis studies on \ac{BH}-star binaries are based on stellar evolution models by \cite{pols1998} \modref{(see, e.g.,\ \citealp{breivik2017,shao2019,olejak2020,chawla2022,shikauchi2023,dicarlo2024})}. In general, the stars in these tracks tend to expand to larger radii compared to the {\sc mist} models.
Most of these works focus on the Galactic population, of which the stellar halo is a negligible component. Hence a direct comparison is not straightforward.
We note that the properties of the few metal-poor stars reported by 
\cite{chawla2022} and \modref{the halo population discussed by  \cite{olejak2020}} seem to be qualitatively consistent with the distribution of the \ac{BH}-star systems \modref{found in our simulations in terms of period and  companion mass}. 

\modref{\cite{dicarlo2024} report a \modlref{BH~star} formation efficiency  for the isolated channel at least one order of magnitude lower than what we found in our simulations. Moreover, these authors claim that any \ac{BH}-star system with an eccentricity larger than 0.5 and a \ac{BH} mass larger than 30 M$_\odot$ are exclusively associated with the dynamical formation channel. 
The origin of this discrepancy is twofold. First, their sample is dominated by the metal-rich population of the Milky Way disc. Secondly, their stellar and binary evolution models are significantly different from the ones adopted in this work. In particular, \cite{dicarlo2024} used stellar tracks by \cite{pols1998} and a different supernova and natal kick model.}

\modref{The number of binaries simulated in our models correspond  to a stellar population with a total mass of $\sim 10^9$ M$_\odot$. This mass is representative of the old and metal-poor Galactic stellar halo  \citep{deason2019}.
However, a single halo realisation suffers from stochastic 
fluctuations. The model fiducial2E8, in which we simulated $2 \times 10^8$ binaries,  indicates that our fiducial model slightly underestimates the number of {\it Gaia}~BH3-like binaries (Table \ref{tab:feff}).  However, by dividing the initial conditions of fiducial2E8 (Sect. \ref{sec:parexp}) into sub-samples of $2 \times 10^7$ binaries, we are able to confirm that the errors quoted in  Table \ref{tab:feff}  realistically capture the uncertainties.}

\subsection{Alternative formation channel} \label{sec:dyn}

{\it Gaia}~BH3 is associated with the ED-2 stream \citep{balbinot2024}. Therefore, it possibly spent a part of its life in a dynamical environment. 
The variety of interactions in dynamical environments has two competitive effects. On the one hand, they can pair \ac{BH}s and stars or create binary systems with orbital properties (e.g. high eccentricities) that boost the production of {\it Gaia}~BH3-like binaries. On the other hand, dynamical interactions could destroy original and newly formed BH-star systems, especially if their orbit is wide and the companion is a low-mass star.

The other two BH-star systems  retrieved from {\it Gaia} data (BH1 \citealt{elbadry2023a,chakrabarti2023}, and BH2 \citealt{elbadry2023b}) 
are difficult to reconcile with isolated binary evolution models \citep{elbadry2023b}. However, they are more easily formed in star clusters (\citealt{rastello2023,dicarlo2024,tanikawa2024}, but see \citealt{kotko2024}).

\cite{marinpina2024} show that BH-star systems dynamically form in an efficient way in globular clusters with mass $10^5-10^6 \ \mathrm{M}_\odot$.  They estimated a formation efficiency of $10^{-5} - 3 \times 10^{-4}$ M$^{-1}_\odot$ by considering all the BH-star systems formed in their Monte Carlo simulations (both escaped or retained in the cluster).  \modref{\cite{dicarlo2024} estimate a formation efficiency of $1.8 \times 10^{-5}$ M$^{-1}_\odot$ by considering star clusters in the mass range $10^{3}-3 \times 10^{4}$ M$_\odot$;  \cite{rastello2023} show that the formation efficiency rises up to $\approx 10^{-3}$ M$^{-1}_\odot$ for low-mass clusters ($\approx10^{2}$--$10^3$ M$_\odot$) at high metallicity ($Z=0.02$).}
Assuming a mass of the ED-2 cluster progenitor in the range $\sim 10^3$--$10^4$ \citep{balbinot2024}, the dynamical channel is one to two orders of magnitude 
more efficient than the isolated channel presented in this work.
However, 
\modref{when considering the mass of the halo \citep{deason2019} and the cluster radial profile (e.g. \citealt{gieles2023}), the difference on the expected formation efficiency of BH-star systems 
in the solar neighbourhood reduces to a factor of three to ten between the isolated and star cluster scenario.}

\modref{
When 
focusing specifically on {\it Gaia}~BH3-like systems, \cite{marinpina2024} find that only 0.6\%  of their BH-star systems \modref{(nine out of 1435)} fall within the same selection box used in our work (Fig. \ref{fig:orb}), and only \modref{two of them (0.1\%)} 
are \modlref{candidates} to be  within an ED2-like stream at the present time (i.e. they experience a late ejection or are still in the cluster at the end of the simulation; Mar\'in Pina, private communication). 
This fraction is a factor of two to ten smaller than what we found in our fiducial model (see Table \ref{tab:feff}) when considering only old {\it Gaia}~BH3-like systems that received a low natal kick (Table \ref{tab:feff}). 
Therefore, the ratio between the dynamical to isolated formation efficiency ranges from 0.6 (isolated channel slightly favoured) to 90 (dynamical channel strongly favoured).
Considering the model with the higher {\it Gaia}~BH3-like formation efficiency (ICthermal; see Table \ref{tab:feff}), the ratio ranges from 0.1 (isolated formation channel favoured) to 20 (dynamical formation channel favoured).
\modref{\citealt{marinpina2024} selected {\it Gaia} BH-3-like systems using a wider eccentricity range: from 0.5 to 0.9. With this selection criterion, the number of {\it Gaia} BH-3-like systems approximately doubles, yet the ratio between the isolated and dynamical channels remains unchanged.}
Taken at face value, these numbers seem to indicate that although the dynamical channel has an overall higher likelihood to form {\it Gaia}~BH3,  we cannot  rule out the possibility that the system is compatible with a primordial binary star that escaped from the ED2 cluster.}

\modlref{A more quantitative comparison and model selection would require a detailed analysis of systematics, including stellar evolution models, selection effects, channel definitions, and halo population models  \modref{(see e.g. \citealt{dicarlo2024})}. Additionally, the use of robust statistical tools is essential, as highlighted by studies such as \citealt{mould2023}. However, these aspects are beyond the scope of this paper.}

\section{Summary}\label{sec:summary}

We have explored the formation of 
{\it Gaia}~BH3 \citep{GaiaBH3}, a non-interacting 
binary system composed of a massive \ac{BH} ($
\approx 30 \ \mathrm{M}_\odot$) and a low-mass giant star ($
\approx 0.8 \ \mathrm{M}_\odot$), by means of population synthesis simulations. 
We have exploited the flexibility of our code {\sc sevn} \citep{iorio2023} to perform the first population study of BH-star systems using the {\sc mist} stellar models \citep{choi2016}. We explored the parameter space 
by considering nine different metallicities, four different supernova kick models, five models for the common envelope, two models for the onset of Roche-lobe overflow, and three different sets of initial conditions. We also implemented a Bayesian sampling analysis to study the distribution of initial conditions that reproduce the observed properties of {\it Gaia}~BH3 exactly. The main takeaways of our analysis are as follows:

\begin{itemize}
\item{Systems such as {\it Gaia}~BH3 represent a common product of isolated binary evolution at 
low metallicity ($Z<0.01$). They form preferentially from 
a massive star (40--60 M$_\odot$) in an initially wide ($P>10^3$ days) and eccentric ($e>0.6$) orbit 
with a low-mass companion. The progenitor binary system never underwent Roche-lobe overflow during its life.}

\item{Our fiducial natal kick model \citep{giacobbomapelli2020} produces low natal kicks \modref{for {\it Gaia}~BH3-like systems}, leaving the initial orbital properties of the binary almost unchanged. The formation of {\it Gaia}~BH3 is possible up to kicks as high as 220 km/s, but \modref{such high kicks require fine-tuning for the properties at BH formation, reducing the overall formation efficiency by about one order of magnitude, and they are not compatible with the association with a cold stellar stream}.}

\item{Focusing on systems older than 10 Gyr, we find a formation efficiency of $\eta_{\rm BHStar}\sim 10^{-6}$ M$_{\odot}^{-1}$ for BH-star binaries in general and $\eta_{\rm {\it Gaia}~BH3-like}\sim 4\times10^{-8}$ M$_{\odot}^{-1}$ for {\it Gaia}~BH3-like systems.}

\item{Considering the Galactic stellar halo model by \cite{deason2019}, we expected up to $\approx$ 4000 BH-star systems born in the stellar halo  from isolated evolution. About \modref{20--120} of them are {\it Gaia}~BH3-like systems in terms of mass and orbital properties. 
}

\end{itemize}

\modref{Overall, {\it Gaia}~BH3 is compatible with a primordial, metal-poor binary system that never evolved via Roche-lobe overflow and experienced a long,  \modlref{`boring'}  life.
Given its association with the ED-2 stream, {\it Gaia}~BH3 was born in a stellar cluster but is
compatible with a primordial binary star that escaped from its parent 
cluster without experiencing significant dynamical interactions.
Unlike {\it Gaia} BH1 and BH2, its orbital properties are easy to explain with current stellar and binary evolution models.  However, uncertainties about massive star evolution (e.g. maximum stellar radius of the BH progenitor) and natal kicks substantially affect our final results. 
Future detections of dormant BHs are thus crucial to set constraints on massive star and natal kick models.}

\section*{Data availability}
Our initial conditions and parameter files are publicly available at 
\href{https://zenodo.org/doi/10.5281/zenodo.11617742}{https://zenodo.org/records/11617742}. The sample of BH-star systems produced in the simulations can be downloaded from \href{https://zenodo.org/doi/10.5281/zenodo.11992265}{https://zenodo.org/records/11992265}, \href{https://zenodo.org/doi/10.5281/zenodo.12188369}{https://zenodo.org/records/12188369}, and \href{https://zenodo.org/doi/10.5281/zenodo.12189734}{https://zenodo.org/records/12189734}.

\begin{acknowledgements}
      \modref{The authors thank the anonymous referee for their suggestions that helped
us to improve the manuscript. }
      We thank Daniel Marín Pina  \modref{and Pasquale Panuzzo} for useful comments, and \modref{Jakub Klencki for his contribution in revising the implementation of the Klencki21 binding energy prescriptions in {\sc sevn}}. 
      GI acknowledges financial support under the National Recovery and Resilience Plan (NRRP), Mission 4, Component 2, Investment 1.4, - Call for tender No. 3138 of 18/12/2021 of Italian Ministry of University and Research funded by the European Union – NextGenerationEU.
      MM, ST, MD, StR, GE, EK, EL and MPV acknowledge financial support from the European Research Council for the ERC Consolidator grant DEMOBLACK, under contract no. 770017. MM, ST, StR and MPV also acknowledge financial support from the German Excellence Strategy via the Heidelberg Cluster of Excellence (EXC 2181 - 390900948) STRUCTURES. MD also acknowledges financial support from Cariparo foundation under grant 55440. AAT acknowledges support from the European Union’s Horizon 2020 and Horizon Europe research and innovation programs under the Marie Sk\l{}odowska-Curie grant agreements no. 847523 and 101103134. 
      SaR acknowledges support from the Beatriu de Pin\'os postdoctoral program under the Ministry of Research and Universities of the Government of Catalonia (Grant Reference No. 2021 BP 00213). MAS acknowledges funding from the European Union’s Horizon 2020 research and innovation programme under the Marie Skłodowska-Curie grant agreement No.~101025436 (project GRACE-BH, PI: Manuel Arca Sedda). MAS acknowledge financial support from the MERAC Foundation.  Part of this work was supported by the German
      \emph{Deut\-sche For\-schungs\-ge\-mein\-schaft, DFG\/} project
      number Ts~17/2--1.
      \modref{Numerical calculations have been made possible through a CINECA-INFN (TEONGRAV) agreement, providing access to resources on Leonardo at CINECA.}
      \modref{This work has made use of data from the European Space Agency (ESA) mission
    {\it Gaia}  (\url{https://www.cosmos.esa.int/gaia}), processed by the {\it Gaia} 
    Data Processing and Analysis Consortium (DPAC,
    \url{https://www.cosmos.esa.int/web/gaia/dpac/consortium}). Funding for the DPAC
    has been provided by national institutions, in particular the institutions
    participating in the {\it Gaia}  Multilateral Agreement.}
\end{acknowledgements}


\bibliographystyle{aa_edited} 
\bibliography{main} 

\end{document}